\documentstyle[aps,prb,preprint,amsmath,graphicx]{revtex}
\tightenlines
\begin{document}
\draft \title{Theoretical Study of Photon Emission from Molecular
Wires}
\author{John Buker and George Kirczenow}
\address{Department of Physics, Simon Fraser University, Burnaby,
British Columbia, Canada, V5A 1S6}
\date{\today}
\maketitle
\begin{abstract}

We explore theoretically the principles that govern photon emission from
single-molecule conductors carrying electric currents between  metallic
contacts. The molecule and contacts are represented by a generic
tight-binding model. The electric current is calculated using Landauer
theory and the photon emission rate is obtained using Fermi's golden rule.
The bias-dependence of the electronic  structure of the molecular wire is
included in the theory in a simple way. Conditions under which significant
photon emission should occur are identified and photon spectra are
calculated. We predict the photon emission rate to be more sensitive than
the electric current to coupling asymmetries between the molecule and
contacts. This has important implications for the design and
interpretation of STM experiments searching for electroluminescence from
individual molecules. We discuss how electroluminescence may be used to
measure important characteristics of the electronic structure of molecular
wires such as the HOMO-LUMO gap and location of the Fermi level of the
contacts relative to the HOMO and LUMO. The feasibility of observing
photon emission from Au/benzene-dithiolate molecular wires is also
discussed.

\end{abstract}
%
\section{Introduction}

A molecular wire is a single molecule (or a few molecules) that forms an
electrically conducting bridge between a pair of metallic nano-contacts.
In recent years molecular wires have been realized in the laboratory
and their electronic transport properties have been
measured\cite{Reed97,Datta97,Metzger,Gim99,Chen,Collier,Reichert}.
There have also been many theoretical advances in modeling such
systems
\cite{Datta97,Samant96,Kemp96,Joach96,Mujic96,Joachim97,%
Yaliraki98,Emberly98,Lang,Emberly99,Yaliraki99,Emberly00,%
Hall00,Mujica,Ventra,Gutierrez,Emberly01,2Emberly01,Damle01%
,Guo,Emb02,Damle02,Damle022,Bratk02,Kornil02}
and the study of transport in molecular wires continues to
be an active area of
experimental and theoretical research\cite{chemphys}.
It has been found that the current flowing through a molecular wire
depends strongly on the electronic structure of the molecule as well
as the geometry of the molecule and contacts and the chemical
bonding between
them\cite{Samant96,Kemp96,Joach96,Mujic96,Joachim97,Yaliraki98,Emberly98%
,Lang,Hall00,Gutierrez,Emberly01,2Emberly01,Damle01,Emb02,Damle02,%
Damle022,Bratk02,Kornil02}.
In a simplified picture, when a potential bias
is applied to the contacts, the
electrochemical potentials of the contacts separate and molecular
orbitals located in the
window of energy between the two electrochemical potentials mediate
electron flow from one
contact to the other. The current rises sharply each time this window of
energy expands to include an additional molecular orbital.

Much work has been done modeling and attempting to understand I-V
characteristics that have been obtained experimentally
\cite{Datta97,Emberly98,Yaliraki99,Hall00,Ventra,2Emberly01,Guo,Bratk02}.
No theoretical studies
have been reported, however, of another potentially important property
of self-assembled molecular wires, namely, photon emission
(electroluminescence) from a
molecular wire carrying an electric current due to electronic
transitions between
molecular orbitals. When a bias voltage is applied to the contacts,
the molecular wire
moves  out of equilibrium, with a flux of electrons passing through
it. Electrons
enter the molecule from the contact with the higher electrochemical
potential, and
drain into the other contact. Since the electrochemical potentials of
the contacts are
no longer the same, one can no longer assume that the orbitals of the
molecule will be
filled up to an energy equal to a common Fermi energy of the
contacts. The passage of
electrons through the molecule may lead to  partial occupation of
various different
molecular orbitals that lie  within the
electrochemical potential window of the
contacts. It is possible that  transitions from one partially
occupied molecular
orbital to another of  lower energy occur, resulting in photon emission.

Although definitive experimental evidence of this effect has not yet been
reported, related phenomena have been observed in scanning  tunneling
microscopy (STM)
experiments:  Systems with an STM tip over
a clean metallic surface are known to emit photons due to the decay of
plasmons\cite{Berndt91}. Recent STM experiments on molecular
monolayers adsorbed on
metal substrates have suggested that a different, molecule-dependent
photon emission
process may also occur: Poirier has  reported molecule-dependent
photon emission in an
STM experiment involving  reduced and oxidized alkanethiol monolayers
adsorbed on
Au(111)\cite{Poirier}.  In an STM experiment involving monolayer
films of C$_{60}$
fullerenes on  Au(110) surfaces, Berndt {\em et al.} observed
enhanced photon
emission when  the STM tip was placed above an individual
molecule\cite{Berndt93}.
Smolyaninov  has reported photon emission at energies similar to
copper phthalocyanine
(CuPc) transition energies in STM experiments involving CuPc adsorbed on
Au\cite{Smolyaninov}. However, in very recent STM experiments by
Hoffmann {\em et al.}
on  hexa-tert-butyl-decacyclene on noble-metal
surfaces\cite{Hoffmann}, it was found
that while the presence of the molecules modulated the photon
emission, the observed
emission appeared to be plasmon-mediated rather than molecular in
character. Thus, the
nature of photon emission mechanisms for STM-molecular
monolayer systems is yet
to be adequately understood.

In all of these STM  experiments, a bias voltage was applied between the
STM tip and metallic  surface, with a molecular monolayer  interposed
between these two contacts as in molecular wire systems. However, there
are  important differences between these STM molecular monolayer  systems
and  self-assembled molecular wires: The coupling between the STM tip  and
the molecule is usually very small compared to the coupling between the
molecule and the substrate, whereas for self-assembled molecular wires
that are bonded chemically to both  metal contacts, the couplings of the
molecule to the source and drain may be comparable  in strength. Also, the
STM experiments involve planar metallic substrates, which are quite
different  from the nanoscopic metallic contacts of some  self-assembled
molecular wires.  It is unclear how these differences would affect photon
emission, so it is of interest to consider photon emission  not only from
STM-molecular monolayer systems but also from more symmetrically coupled
self-assembled molecular wires.

Electroluminescence from molecular wires is a potentially important effect,
not only because of its intrinsic fundamental interest, but also as a
novel experimental probe of molecular nano-electronic devices. For
example, as we will show  below, by observing the photon emission spectrum
of a molecular wire, important information  about its the electronic
structure, including the energies of the molecular orbitals, their
locations relative to the Fermi level, and their  occupations as a
function of bias voltage, may in principle be obtained. There is
currently disagreement between different theoretical models of 1,4
benzene-dithiolate (BDT) attached to gold
contacts\cite{Emberly98,Hall00,Ventra,Damle01,Damle02,Bratk02} regarding
the energies  of the highest occupied molecular orbital (HOMO) and the
lowest unoccupied  molecular orbital (LUMO), relative to the Fermi level
of  the contacts. If the photon emission  spectra of such systems were
modelled, and compared with experiment,  important new insights into the
most appropriate models for molecular wires could be obtained.

In this article, we present and solve a generic model for photon  emission
from a molecular wire. As this is the first theoretical study of this
effect, our purpose will be to develop a basic qualitative picture of the
underlying mechanism, and to examine how adjustment of the various model
parameters, to reflect different experimental situations, should affect
photon emission. We use a one-dimensional tight-binding model that
attempts to capture the important physics involved in a way that is simple
and intuitively reasonable, and is qualitatively consistent with what is
already known about molecular wire electronic structure and transport.
Another objective of this work is  to provide some guidance to
experimentalists as to which types of  systems to study  in order to
observe photon emission. We find that, under  certain circumstances, the
model predicts that photon emission should occur, due  to molecular
orbital transitions.  We show how molecular orbital occupations  behave as
a function of applied bias  voltage, and how varying the model  parameters
changes this behaviour. Fermi's Golden  Rule is used to calculate photon
emission spectra. We show how the emission depends on molecular orbital
occupations, and the locations of the molecular orbitals relative to the
Fermi levels of the contacts.

Our model and theoretical approach are described in Section II. The
results of our calculations and their interpretation are presented in
Section III.  We conclude in Section IV by summarizing our results, and
discussing their  implications for possible experiments directed at
observing photon emission from BDT attached  to gold contacts.

\section{The Model}

Each metal contact will be modelled as a one-dimensional
tight-binding chain. The
molecular wire is modelled as a pair of atoms placed next to the
origin, forming
a bridge between the two contacts (see Fig. \ref{fig1}). The model
Hamiltonian of this system is
\begin{eqnarray}
H = \sum_{n<0}\epsilon _L |n\rangle \langle n|
+ \beta(|n-1\rangle \langle n|
+ |n\rangle \langle n-1|) \nonumber \\ + \sum_{n>0}\epsilon _R |n\rangle
\langle n| + \beta(|n+1\rangle \langle n| + |n\rangle \langle n+1|)
\nonumber \\ + \epsilon _a|a\rangle \langle a| + \epsilon _b|b\rangle
\langle b| + \beta_{-1,a} (|-1\rangle \langle a| + |a\rangle \langle -1|)
\nonumber \\ + \beta_{1,b} (|1\rangle \langle b| + |b\rangle \langle 1|)
+ \beta_{a,b} (|a\rangle \langle b| + |b\rangle \langle a|)
\label{hamil}
\end{eqnarray}
where $\epsilon_L$ and $\epsilon_R$ are the site energies of the left
(source) and right (drain) contacts, and $\epsilon_a$ and
$\epsilon_b$ are the site energies of atoms $a$ and $b$ of the molecule,
which depend on the bias voltage applied to the wire and the nature of the
metal contacts, as well as the identities of atoms
$a$ and $b$. $\beta$, $\beta_{-1,a}$, $\beta_{1,b}$ and $\beta_{a,b}$  are
the hopping amplitudes between atoms in the contacts, between atom a of
the molecule and the left contact, between atom b and the right contact,
and between atom a and atom b, respectively. $|n\rangle$ represents the
orbital of an atom in one of the contacts, and $|a\rangle$ and $|b\rangle$
represent the atomic orbitals of the molecule. We consider the orbitals of
different atoms to be orthogonal. However, this model may easily be
extended to systems where this is not the case\cite{Emberly99}. We take
the the electrochemical potentials of the source and drain  contacts to be
$\mu_S = E_F+eV_{bias}/2$ and $\mu_D = E_F - eV_{bias}/2$ where $V_{bias}$
is the bias voltage applied between  them and $E_F$ is their common Fermi
level at zero applied bias. The applied bias also  affects the site
energies $\epsilon_L$ and $\epsilon_R$ of the contacts so that
$\epsilon_L =
\epsilon_{contacts} + eV_{bias}/2$ and
$\epsilon_R = \epsilon_{contacts} - eV_{bias}/2$, where
$\epsilon_{contacts}$ is the zero bias site energy. Our treatment of the
effect of the bias voltage on site energies $\epsilon_a$ and
$\epsilon_b$ of the molecule itself is described at the end of Section II.

Electrons exist in the form of Bloch waves in the contacts, and
undergo reflection or transmission when they encounter the two-atom
molecule. Their wavefunctions are of the form
\begin{equation}
|\psi\rangle = \sum_{n<0}(e^{iknd} + re^{-iknd}) |n\rangle +
\sum_{n>0}te^{ik^{\prime}nd} |n\rangle + c_a |a\rangle + c_b |b\rangle
\label{psi}
\end{equation}
where $d$ is the lattice spacing, and $t$ and $r$ are the transmission and
reflection coefficients. The electrons have eigenenergies of the form $E =
\epsilon_i + 2\beta cos(kd)$. This equation holds for both $\epsilon_i =
\epsilon_L$ and $\epsilon_i = \epsilon_R$, so when an electron with initial
wavevector $k$ undergoes transmission, its wavevector changes (to
$k^\prime$) due to the difference between $\epsilon_L$ and $\epsilon_R$.
For certain $k$, there are no real solutions for $k^{\prime}$. In these
cases,
$k^{\prime}$ becomes complex, and the transmitted Bloch wave is
evanescent. By applying $\langle -1|$, $\langle a|$, $\langle b|$, and
$\langle 1|$ to $H|\psi\rangle$, analytic expressions for the transmission
and reflection coefficients are obtained:
\begin{eqnarray}
t = \frac{2i\beta_{-1,a}\beta_{1,b}\beta_{a,b}\beta
sin(kd)}{\beta_{a,b}^2 \beta^2 -
[\beta(E - \epsilon_b) - \beta_{1,b}^2 e^{ik^\prime d}][\beta(E
- \epsilon_a) -
\beta_{-1,a}^2 e^{ikd}]} \\
r = - \frac{[\beta(E-\epsilon_a) - \beta_{-1,a}^2
e^{-ikd}][\beta(E-\epsilon_b) -
\beta_{1,b}^2 e^{ik^\prime d}] - \beta_{a,b}^2 \beta^2}{[\beta(E -
\epsilon_a) -
\beta_{-1,a}^2 e^{ikd}][\beta(E - \epsilon_b) - \beta_{1,b}^2
e^{ik^\prime d}] -
\beta_{a,b}^2 \beta^2}
\label{tr}
\end{eqnarray}
The coefficients $c_a$ and $c_b$ for the atomic orbitals in the molecule
are similarly obtained, in terms of the transmission and reflection
coefficients:
\begin{equation}
c_a = \frac{\beta(1+r)}{\beta_{-1,a}}, \quad c_b =
\frac{\beta t}{\beta_{1,b}}
\label{cacb}
\end{equation}
The transmission probability is given by
\begin{equation}
T = |t|^2 \frac{v(k^{\prime})}{v(k)} =
|t|^2 \frac{sin(k^\prime d)}{sin(kd)}
\label{T}
\end{equation}
where $v(k)$ is the velocity of the incoming wave and $v(k^\prime)$ is the
velocity of the transmitted wave. In our model, $T$ depends on the energy
of the incoming electron, as well as on $V_{bias}$. The $V_{bias}$
dependence arises from the effect of $V_{bias}$ on the site energies of
the  contacts (described above) and of the molecule that will be discussed
below. Using Landauer theory\cite{Lan57}, an expression for the current is
obtained:
\begin{equation} I = \frac{2e}{h}\int_{\mu_D}^{\mu_S}T(E,V_{bias})dE
\label{current}
\end{equation} Here we assume the temperature to be 0 K, so the Fermi
functions are trivial, and the transmission probability is only integrated
through the Fermi energy window.

To calculate emission spectra for the molecular wire we
use the expression for the spontaneous emission rate of a system emitting
photons into empty space, using Fermi's Golden Rule\cite{Ballentine}.
The emission rate is given by
\begin{equation}
\frac{4e^2\omega ^3}{3\hbar c^3} |\langle\psi_f|{\bf x}|\psi_i\rangle|^2
\label{emission}
\end{equation}
where $\psi_i$ and $\psi_f$ represent initial and final states, and
$\hbar \omega$ is their difference in energy. We consider emission only
from the molecular sites $a$ and $b$. The emission rate is
therefore approximated by
\begin{equation}
R = \frac{4e^2\omega ^3}{3\hbar c^3} |c_{a,f}^*c_{a,i}\langle a|
{\bf x}|a\rangle + c_{b,f}^*c_{b,i}\langle b|{\bf x}|b\rangle|^2
\label{molemit}
\end{equation}
where $i$ and $f$ label initial and final states. The overlap terms
$\langle a|{\bf x}|b\rangle$ and $\langle b|{\bf x}|a\rangle$ are
neglected since they should be small compared to
$\langle a|{\bf x}|a\rangle$ and $\langle b|{\bf x}|b\rangle$. We
approximate $\langle a|{\bf x}|a\rangle$ and $\langle b|{\bf x}|b\rangle$
by the locations of their atomic centers, so that
$\langle a|{\bf x}|a\rangle =
-\frac{b}{2}$, $\langle b|{\bf x}|b\rangle =
\frac{b}{2}$ ($b$ being the molecular bond length). Thus, we have
\begin{equation}
R(k_i,\omega) = \frac{e^2\omega ^3 b^2}{3\hbar c^3} |c_{b,f}^*c_{b,i} -
c_{a,f}^*c_{a,i}|^2
\label{molemit2}
\end{equation}
To calculate the emission rate as a function of photon energy, we must
consider all electron states of the system, incoming from both the source
and drain contacts. Since we assume the temperature to be 0 K, all states
up to the electrochemical potential of the appropriate contact are
occupied. $\psi_f$ must be initially unoccupied, and it must be of lower
energy than $\psi_i$. Therefore, we consider transitions from occupied
initial states that are incoming from the source, to final states, within
the electrochemical potential window, that are incoming from the drain.
After normalizing the wavefunctions and converting the sum over k-states
(and spin) into an integral over energy, an expression for the photon
emission spectrum (for a given bias voltage) is obtained:
\begin{equation}
f(\omega) = \frac{1}{2 \pi} \int_{\mu_D+\hbar\omega}^{\mu_S}
\frac{R(k_i,\omega)}
{-\beta sin(k_i d)}dE_i
\label{spectrum}
\end{equation}

In order to interpret our results physically, it is useful to relate the
emission spectra to total occupations of the molecular orbitals, as a
function of bias voltage. For this purpose, we project eigenstates of the
system given by equation~(\ref{psi}) onto the bonding and antibonding
orbitals of the isolated molecule, whose coefficients will be labelled
$c_O$ and $c_{O^*}$ respectively. Their total occupations are calculated
by summing over all occupied electron states (including spin) incoming
from both contacts. This sum is converted into an integral, and an
expression for the total bonding and antibonding orbital occupations
($O,O^*$) is obtained:
\begin{equation}
(O,O^*) = \frac{1}{2 \pi} \sum_{contacts \hspace{0.1cm} i} \int
\frac{|c_O,c_{O^*}|^2}{-\beta sin(k_i d)} dE_i
\label{occupy}
\end{equation}
Changes in the occupations of the bonding and antibonding orbitals in
response to changes of the applied bias voltage may result in some
charging of the molecule.  If this occurs, the  charging causes an
electrostatic shift of the molecular energy levels   that in turn severely
limits the actual charging that takes place\cite{Emberly01}. In the
present work we approximate the shift of the molecular levels in response
to the  applied bias by adjusting $\epsilon_a$ and $\epsilon_b$ equally so
as to maintain the net charge that the molecule has at zero bias.  The
simplicity of this approximation is in keeping with the generic nature of
the molecular  wire model that we consider here. It yields behavior of the
molecular levels with bias that is physically reasonable, and as will be
seen in Section  III B, remarkably similar to that obtained  from {\em ab
initio} calculations for some molecular wire systems.

In the remainder of this article we will consider the case where the
Fermi level of the leads at zero bias falls within the HOMO-LUMO gap of
the molecule, as is typical of molecular wires such as the
Au/benzene-dithiolate/Au system. Thus we will identify the bonding and
antibonding orbitals $O$ and $O^*$ with the HOMO and LUMO,
respectively, of the isolated molecule. (The terms $O$ and HOMO will
henceforth be used interchangeably as will $O^*$ and LUMO.) Our model
parameters will be chosen so that at zero bias the Fermi level of the
leads lies between
$O$ and $O^*$. As will be seen below, if the Fermi level is
well above $O$ and well below $O^*$ (but $O$ and $O^*$ are within the
electronic energy band of the leads) it follows from Eq. 12 that at zero
bias $O$ is almost completely filled with electrons and $O^*$ is almost
empty, although $O$ is never {\em totally} full and $O^*$ is never
completely empty because of the level broadening that occurs due to
hybridization between the states of the molecule and leads. In this way
the HOMO and LUMO character of the
$O$ and $O^*$ orbitals is expressed within our model when
the molecule couples to the leads. By choosing different values of the
site energies
$\epsilon_a$ and $\epsilon_b$ at zero bias, the energies and occupations of
the $O$ and $O^*$ orbitals can be varied. Thus the effects of differing
amounts of charge transfer between the molecule and leads can also be
studied within our model.

We also note in passing that although our results are
described here in terms of electron transport and
optical transitions between occupied and empty electron states, an
equivalent description can be given in terms of
electron and hole transport and optical transitions.

\section{Results}
We consider the case where the energy bands of the contacts are partially
filled ($k_F = \frac{\pi}{2d}$). In numerical results presented below the
hopping parameters are chosen to have values typical of molecular wire
systems ($\beta = -5$ eV,
$\beta_{a,b} = -1.5$ eV), resulting in a HOMO LUMO gap of 3 eV for the
isolated molecule.

\subsection{Symmetric molecule-contact couplings; HOMO LUMO gap
centred at $E_F$.}
In this section, The HOMO-LUMO gap of the molecule is chosen to be centred
at the zero bias Fermi level of the contacts, and the molecule-contact
couplings have equal values. First, we consider the case where these
couplings have values similar to those in chemically bonded molecular wire
systems ($\beta_{-1,a}=\beta_{1,b}=-1.0$ eV). Fig. \ref{fig2}a shows how
the source and drain electrochemical  potentials $\mu_S$ and $\mu_D$ and
the energies of the bonding ($O$) and  antibonding  ($O^*$) orbitals
behave as a bias is applied to the  contacts. At zero bias, $O$ is located
at -1.5 eV and
$O^*$ at +1.5 eV. Fig. \ref{fig2}b shows the electron transmission
probability through the molecule at zero bias obtained from eq. \ref{T}.
The bonding and antibonding orbitals provide channels for electron
transmission, so there are peaks at electron energies of -1.5 eV and +1.5
eV. The  transmission peaks are somewhat broadened due to the coupling  of
the molecule to the contacts, which causes the discrete molecular
orbitals ($O$ and $O^*$) to  hybridize with the continuum of states in the
contacts.  Returning to Fig.
\ref{fig2}a, as $V_{bias}$ increases, the electrochemical potentials  of
the source ($\mu_S$) and drain ($\mu_D$) separate. At $V_{bias}=3$ V,
$\mu_S$ moves above $O^*$ and $\mu_D$ moves below $O$. In this symmetric
case, this does not result in any change in the energies of $O$ and $O^*$.
Fig. \ref{fig2}c shows the electron occupation of the molecular wire, as
projected onto the bonding ($O$) and antibonding ($O^*$) orbitals of the
isolated molecule. At zero bias, $O$ is almost fully occupied, and $O^*$
is almost empty. As $V_{bias}$ increases, $O$ partially empties. $\mu_D$
moves below $O$ in energy, so electrons from the drain no longer
contribute to the filling of the bonding orbital, while electrons from the
source continue to contribute, so $O$ becomes half-filled. Similarly,
electrons from the source start to fill
$O^*$ as $\mu_S$ approaches $O^*$ in energy, and finally $O^*$ becomes
half-filled as well. As $O$ and $O^*$ empty and fill with $V_{bias}$, the
total occupation remains constant (two electrons), due to the symmetric
placement of $E_F$, at the centre of the HOMO LUMO gap. Therefore, there
is no tendency for the molecule to charge, and the orbital energies remain
constant (Fig. \ref{fig2}a). Notice that the filling and emptying of $O$
and $O^*$, as a function of $V_{bias}$, is gradual. This is due to the
fact that the continuous density of states associated with the broadened
resonances in Fig. 2b enters the electrochemical potential window between
$\mu_S$ and $\mu_D$ gradually. Fig. \ref{fig2}d  shows the orbital
occupation for the case of weak molecule-contact coupling  ($\beta_{-1,a}
=\beta_{1,b}=-0.2$ eV). The  weak coupling results in a more  sharply
peaked molecular density of states, so $O$  and $O^*$ empty and fill much
more abruptly. Fig. \ref{fig2}e shows the resulting total photon emission
rate for  (i)
$\beta_{-1,a} =\beta_{1,b}=-0.2$ eV and (ii)
$\beta_{-1,a}=\beta_{1,b}=-1.0$  eV. Emission is strong when the energies
of $O$ and $O^*$ are inside the  electrochemical potential window. In case
(i), emission increases more rapidly with  bias near
$V_{bias}=3.0$ V than in case (ii), because $O$ and $O^*$ empty  and fill
more abruptly. Fig.
\ref{fig2}f shows how, within our model,  the emission spectrum for  case
(ii) changes with bias. As expected,  emission is peaked around the
transition energy of the molecule (3 eV)  at higher bias. (Weaker
molecule-contact couplings  would result in more sharply peaked emission
spectra, due to weaker hybridization and  broadening of molecular
orbitals.) Since photons cannot be emitted at energies higher than
$eV_{bias}$ the spectra cut off at this energy. This is one of the factors
that contributes to the noticeable shift of the emission peak upwards in
energy with increasing
$V_{bias}$. Another factor contributing to this blue shift is the  cubic
dependence of the emission rate on the photon energy $\hbar \omega$
   in equation~(\ref{molemit}). This explains the significant photon
emission seen in Fig. \ref{fig2}f for $V_{bias}=6$ V at energies well
above the molecule's nominal HOMO-LUMO gap energy, due to transitions from
the higher  energy tail of $O^*$, and to the lower energy tail of $O$.

\subsection{Symmetric molecule-contact couplings;
$E_F$ close to HOMO or LUMO.}
In many experimental situations, the HOMO-LUMO gap is not centred at the
Fermi level of the contacts. Often, either the HOMO or LUMO is close to
$E_F$\cite{Ventra,Damle01,Damle02}. In Fig. \ref{fig3}a, $O$ has a zero
bias energy located just below $E_F$. In our model, as $V_{bias}$
increases,
$O$ and $O^*$ first decrease in energy following $\mu_D$. Then, at around
3  V, as the energy of
$O^*$ approaches $\mu_S$, $O$ and $O^*$ begin to rise in energy with
$\mu_S$. The opposite situation occurs when $O^*$ has a zero bias value
just above $E_F$ (Fig. \ref{fig3}b). In this case, $O$ and $O^*$ first
rise in energy with
$\mu_S$, then fall with $\mu_D$ after $V_{bias} = 3$ V. This behaviour is
nearly identical to recent results obtained through self-consistent
density-functional calculations for the HOMO and LUMO energies of a gold
nanowire attached to two gold contacts\cite{Damle022}, suggesting that our
model's approach to charging can be quite a good approximation for some
molecular wires. Fig. \ref{fig3}c shows the orbital  occupations for the
case of $O$ just below $E_F$ in energy  at zero bias (same situation as
Fig. \ref{fig3}a). The total occupation is less than 2 electrons (1.6
electrons) due to the fact that, at zero bias, $O$ is very close to $E_F$,
so that its high energy tail is above $E_F$ and is unoccupied. As $\mu_D$
decreases, the molecular orbitals drop in energy so that the total
molecular charge is maintained at its zero bias value. Eventually, as
$\mu_S$ increases, it approaches the antibonding energy, causing $O^*$ to
start to fill. The orbitals stop dropping in energy, so that any filling
of $O^*$ is accompanied by an emptying of $O$. Once $O$ is significantly
above $\mu_D$, its occupation becomes constant with further changes in
bias, at about 1 electron, as only the source contact contributes
electrons. Both orbitals now move upwards in energy, causing $O^*$ to stop
filling so that the total charge is still maintained at its zero bias
value. The high energy tail of $O^*$ stays above $\mu_S$, so that the
orbital only fills to about 0.6 electrons. Fig. \ref{fig3}d similarly
shows the orbital occupations for the case of
$O^*$ just above $E_F$ at zero bias (the situation in Fig.
\ref{fig3}b). In this case, the  total occupation is more than 2 electrons
(2.4 electrons) due to  the fact that
$O^*$ is partially occupied at zero bias, because its low energy tail is
below
$E_F$. $O$ partially empties and $O^*$ partially fills as the orbitals
enter the electrochemical potential window at a bias voltage near 3 V. The
occupation of
$O^*$ becomes about  1 electron, coming from the source contact. The
occupation of $O$ is greater  (1.4 electrons) because the low energy tail
of $O$ is  still below $\mu_D$. Fig. \ref{fig3}e shows the total emission
rate predicted by our model for the case where $O$ is just below $E_F$ at
zero bias. A nearly identical  emission rate is predicted for the case of
$O^*$ just above $E_F$. In both cases,  emission becomes strong above
$V_{bias}=3$ V. This is the point at which $O$ and $O^*$ enter the
electrochemical potential window, and therefore are both partially
occupied. Emission at high bias is less than half as strong as in the case
of HOMO and  LUMO that are symmetric about $E_F$ (Fig. \ref{fig2}), due to
the fact that either
$O^*$ is not as highly occupied (Fig. \ref{fig3}c) or $O$ is not as empty
(Fig.
\ref{fig3}d),  as in Fig. \ref{fig2}c. Fig. \ref{fig3}f shows the emission
spectrum for  the  case of $O$ just below $E_F$ at zero bias, for
$V_{bias}=6$ V. (A very similar spectrum is predicted for the case of
$O^*$ just above $E_F$.) The spectrum is  not blue-shifted  as strongly as
in Fig. \ref{fig2}f, because the higher energy tail of
$O^*$ remains unoccupied (Fig.
\ref{fig3}a) (or the lower energy tail of $O$ remains fully occupied (Fig.
\ref{fig3}b)), resulting in fewer high energy transitions.

\subsection{Asymmetric molecule-contact couplings; HOMO-LUMO gap
centred at $E_F$.}
By examining the effects of varying $\beta_{-1,a}$ and $\beta_{1,b}$  we
are also able explore the behavior of photon emission from molecular wires
with asymmetric molecule-contact couplings, such as systems where one of
the contacts  is an STM tip. In this section, we consider the case where
the HOMO-LUMO gap of the  molecule is centred at the zero bias Fermi level
of the contacts. First, we consider highly asymmetric couplings:
$\beta_{-1,a}=-0.1$ eV, $\beta_{1,b}=-1.0$ eV. Fig. \ref{fig4} shows how
$O$ and $O^*$ change in energy as a function of bias. The coupling of the
molecule to the drain is much greater than to the source, so the drain has
a much greater effect on orbital occupations. As
$V_{bias}$ increases from 0 V, the energy levels of $O$ and $O^*$ drop at
the same rate as $\mu_D$. In this way the occupations of $O$ (although a
part of its high energy tail is above $\mu_D$) and $O^*$ (due to the part
of its low energy tail that is below $\mu_D$) change little, and the total
charge is  maintained at its zero bias value. As $O^*$ crosses $\mu_S$ in
energy, its occupation increases very slightly. This causes a slight
deviation of the orbital energies from a straight line, as $O$ empties
slightly by moving closer in energy to $\mu_D$. The occupation of $O$
remains close to 2 electrons throughout, and $O^*$ remains almost
completely unoccupied. In agreement with this, our model predicts that
photon emission is extremely weak in this situation. If the values of
$\beta_{-1,a}$ and $\beta_{1,b}$ are switched, $O$ and $O^*$ rise in
energy with $\mu_S$ instead. The orbital occupations are similar to those
in the opposite situation, since this time the source has a much greater
effect on occupations. So, photon emission in this reverse situation is
also negligible. This lack of significant photon emission from the
molecule, caused by the asymmetry of the contact couplings, is a possible
explanation for the lack of molecular-based  photon emission observed in
recent STM-HBDC monolayer experiments by Hoffmann {\em et
al.}\cite{Hoffmann}.

To understand better the dependence of the photon emission on the
asymmetry  of the couplings, it is useful to consider an intermediate case
of asymmetric couplings shown in Fig. \ref{fig5}: $\beta_{-1,a}=-0.6$ eV,
$\beta_{1,b}=-1.0$ eV. Fig. \ref{fig5}a shows how the model predicts
$O$ and $O^*$ to change in energy as a function of bias. Up to
$V_{bias}=3$ V, $O$ and $O^*$ change little in energy. Above
$V_{bias}=3$ V, both $O$ and $O^*$ descend in energy with $\mu_D$. Fig.
\ref{fig5}b shows the orbital occupations. For $V_{bias} < 3$ V, $O^*$
fills significantly as it approaches $\mu_S$. For this reason, at
$V_{bias} < 3$ V, $O$ and $O^*$ do not drop much in energy, as they do in
Fig. \ref{fig4} due to the emptying of the high energy tail of $O$. At
$V_{bias}=3$ V, $O^*$ partially fills as it crosses $\mu_S$, but the drain
couples more strongly to the molecule. Therefore, $O$ and $O^*$ are pulled
down with $\mu_D$, such that $O$ only empties to the extent that $O^*$
fills, so that the total charge is maintained at its zero bias level. The
result is that $O$ empties to about 1.5 electrons, and $O^*$ fills to 0.5
electrons. Fig. \ref{fig5}c shows photon emission for this case, and other
intermediate cases. For this case ($\beta_{-1,a}=-0.6$ eV), emission is
quite drastically reduced compared to the case of symmetric couplings, but
is not negligible. For the reverse situation ($\beta_{-1,a}=-1.0$ eV,
$\beta_{1,b}=-0.6$ eV), the reverse arguments apply, and the result is
that the emission is similar. The degree of contact coupling asymmetry
clearly plays a key role in determining the strength of photon emission.
Our results indicate that in order for an STM experiment to detect
molecular-based photon emission, the STM tip must not be very weakly
coupled with the molecule. Since coupling strength varies exponentially
with coupling distance, any molecular-based photon emission observed in an
STM experiment will be very sensitive to the tip-sample distance. If the
tip is too far from the sample, emission will be negligible.

In order to facilitate comparison of this prediction with the results  of
potential experiments,  it is useful to relate contact coupling asymmetry
to  the amount of current  flowing through a molecular wire as well,
because, unlike coupling strength, current is a directly measurable
quantity.  Fig.
\ref{fig5}d shows how the current (obtained from eq.~(\ref{current}))
changes with
$V_{bias}$ for various values of $\beta_{-1,a}$, holding $\beta_{1,b}$
fixed at $-1.0$ eV. As $O$ and $O^*$ enter the Fermi energy window, these
orbitals act as channels for electron transmission, and the current
increases as a result. Asymmetric contact couplings result in reduced
current flow. Notice, however, that the current flow is not affected  by
strongly asymmetric couplings as much as is the photon emission rate. Fig.
\ref{fig5}e compares how photon emission and current depend on contact
coupling asymmetry, for $V_{bias}=6$ V. Emission has a power law
dependence (roughly ${\beta_{-1,a}}^{3.8}$), while the  dependence of the
current on the asymmetry is clearly much less strong. Fig.
\ref{fig5}f shows the calculated photon  emission vs. the current, for
$V_{bias}=6$ V. At weak currents, the photon emission rapidly becomes
negligible. This is an important consideration to keep in mind when
attempting to observe molecular photon emission experimentally.

\subsection{Asymmetric molecule-contact couplings; $E_F$ close to
HOMO or LUMO.}
Our results suggest that, for a HOMO-LUMO gap that is symmetric about
$E_F$, asymmetric molecule-contact couplings, such as those where one of
the contacts is an STM tip, result in very low photon emission rates. We
consider now strongly asymmetric couplings ($\beta_{-1,a}=-0.1$ eV,
$\beta_{1,b}=-1.0$ eV) for the case of HOMO or LUMO very close to $E_F$ at
zero bias. This coupling configuration is  analogous to an STM experiment
performed under forward bias (the STM tip acts as the  (weakly coupled)
source, from which electrons flow). The opposite configuration
($\beta_{-1,a}=-1.0$ eV, $\beta_{1,b}=-0.1$ eV) is analogous to an  STM
experiment performed under reverse bias (the STM tip acts as the (weakly
coupled) drain). Fig. \ref{fig6}a shows how, for the forward bias
situation, $O$ and
$O^*$ change in energy as a function of bias, for the case of $O$ very
close to $E_F$ at
$V_{bias}=0$. $O$ and $O^*$ drop in energy following the  electrochemical
potential
$\mu_D$ of the strongly  coupled contact. Even as $O^*$ crosses
$\mu_S$, the orbital energies continue to decrease almost linearly
because, unlike in Fig.
\ref{fig4}, $O$ is very  close to $\mu_D$, so that the density of states
at $\mu_D$ is high and the weak filling of $O^*$ that occurs as it crosses
$\mu_S$ can be compensated by a weak emptying of $O$ with only a very
small deviation of
$O$ from its linear behavior. Fig.
\ref{fig6}b shows the energies of
$O$ and $O^*$ for the case of reversed contact coupling strengths
($\beta_{-1,a}=-1.0$ eV, $\beta_{1,b}=-0.1$ eV), which is equivalent  to
performing an STM experiment under reverse bias. This time, $O$ and $O^*$
rise  in energy following the electrochemical potential $\mu_S$. $O^*$
never approaches
$\mu_S$, so it remains empty.  Fig. \ref{fig6}c,d show photon emission for
the forward bias situation (Fig. \ref{fig6}a). Biases above 3 V result in
a small amount of  photon emission because, unlike in the case of Fig.
\ref{fig4}, $O$ is only partially  occupied, so some transitions from
$O^*$ may occur. Emission is still weak (1-2 orders of magnitude less than
for symmetric contacts), because $O^*$ becomes  only weakly occupied when
crossing the Fermi level of the weakly coupled contact $\mu_S$.  The
emission spectrum, shown in Fig.
\ref{fig6}d, is peaked at an energy slightly  less than 3 eV, because  the
lower energy tail of $O$ is completely filled, and may not receive
transitions. In  the reversed coupling situation  (Fig. \ref{fig6}b),
photon emission is  negligible, since $O^*$ never approaches
$\mu_S$ and remains empty. Now we consider the cases where $O^*$ is very
close to $E_F$ at $V_{bias}=0$ (Fig. \ref{fig6}e,f). Again, the energies of
$O$ and $O^*$ follow the Fermi level of the strongly coupled contact.
Therefore, for the case of weak source contact coupling (Fig. \ref{fig6}e),
there is negligible emission because $O$ never crosses $\mu_D$ and remains
completely filled. For the case of weak drain contact coupling (Fig.
\ref{fig6}f), there is (weak) photon emission, because $O$ enters the
electrochemical potential window and (slightly) empties. (The emission
curves for  this case are very similar to Fig.
\ref{fig6}c,d.) This property of asymmetrically coupled  systems may  be
used to help distinguish a system where the HOMO is close to
$E_F$ from a system where the LUMO is close to $E_F$, with the use of an
STM as the weakly coupled contact. Within our model, if the HOMO is close
to
$E_F$, the result is photon emission under high forward bias and negligible
emission under reverse bias, whereas if the LUMO is close to $E_F$, the
result is emission under high reverse bias and negligible emission under
forward bias. The implications of this may be important for molecular
electronics research. For example, there is currently disagreement over
whether $E_F$ is located close to the HOMO or the LUMO in the system of
BDT attached to gold
contacts\cite{Emberly98,Hall00,Ventra,Damle01,Damle02,Bratk02}. The
photon  emission characteristics of such a system could help determine the
locations of  the HOMO and LUMO, and therefore which type of model is more
appropriate for such a system.

\section{Conclusions}

We have examined theoretically the possibility of photon emission occuring
due to molecular transitions in current-carrying molecular wires. Our
results suggest that significant photon  emission may occur for bias
voltages near or above the HOMO-LUMO energy gap of the molecule, depending
on the  specifics of the electronic structure of the molecule and contacts
and of the  couplings between the molecule and the source and drain. The
predicted photon spectra are peaked near the gap energy\cite{exciton}. If
two molecular orbitals are located in the energy window between the
electrochemical potentials of the contacts, they will  both be partially
occupied and, as long as transitions between the orbitals are not
forbidden, transitions from the higher energy orbital to the lower energy
orbital should occur, resulting in photon emission. We find that varying
the strength of the molecule-contact couplings changes the emission
strength and  spectrum.  Weak, symmetric couplings result in a sharply
peaked emission  spectrum. Asymmetric  couplings result in reduced photon
emission.  Emission from strongly asymmetrically  coupled systems (such
as  some STM-molecular monolayer systems) is predicted to be  extremely
weak if the zero bias value Fermi level $E_F$ of the contacts is not close
to the  HOMO or LUMO energy of  the molecule. If $E_F$ is close to the
HOMO or LUMO energy, the amount of photon emission observed under forward
bias and under reverse bias, in an STM experiment, may be compared in
order to determine which of the orbitals is close in energy to $E_F$.

Whether $E_F$ is close to the HOMO or LUMO energy is a difficult
unresolved issue in current research on molecular wires such as BDT
attached to gold contacts; {\em ab initio} calculations based on
different models and approximations
\cite{Ventra,Damle01,Damle02} have yielded opposite results in this
regard.  It is believed that the HOMO and LUMO  of BDT attached to gold
contacts are composed  of bonding and antibonding orbitals ($\pi$ and
$\pi^*$). The present work suggests  that, if the contact couplings are
symmetric, photon emission  will not be  drastically different whether
$E_F$ is close to the HOMO or to the LUMO: Each orbital will  become
partially occupied as the bias voltage becomes greater than the  HOMO-LUMO
gap,  and, so long as
$\pi^*\-->\pi$ transitions are allowed,  this will result in photon
emission. Optical transitions are forbidden if
$\langle\psi_f|{\bf x}|\psi_i\rangle = 0$. We have used extended
H\"{u}ckel theory to examine the HOMO and LUMO of BDT\cite{Yaehmop}. The
contacts were not included in this calculation. By symmetry analysis of
the HOMO and LUMO, we found that
$\langle HOMO|y|LUMO\rangle\not= 0$, where $y$ is in the direction of the
plane of the molecule, perpendicular to the contacts. Therefore, optical
transitions should be allowed, and would result in the emission of
y-polarized photons. Experimental evidence of BDT optical transitions is
available in the form of absorption spectra for BDT in hexane\cite{Koba}.
Absorption occurs for energies higher than about 4 eV, indicating a
possible experimental value for BDT's HOMO-LUMO energy gap. Transport
measurements on Au/BDT/Au molecular wires have been  reported at bias
voltages as high as approximately 5V\cite{Reed97}. Thus it appears
to  be quite reasonable to undertake an experimental search for
electroluminescence due to  molecular transitions in Au/BDT/Au systems. An
experimental determination of the HOMO-LUMO gap of BDT in the presence of
the Au contacts by this means would be  an important test of the validity
of various theories of molecular wires. If one of the  Au contacts is an
STM tip, electroluminescence measurements may also be expected to be
helpful in determining the location of the Fermi level of Au relative to
the  HOMO and LUMO states of the molecular wire.

The study of electroluminescence from molecular wires has only just begun.
We hope that the theoretical ideas presented here will stimulate further
experimental and theoretical work on this interesting and potentially
important topic.

This work was supported by NSERC and by the Canadian Institute for
Advanced Research.

\begin{figure}[!t]
\caption{A schematic diagram of the model molecular wire. The source
and drain contacts are semi-infinite. The two atoms in the center
represent the molecule.}
\label{fig1}
\end{figure}

\begin{figure}[!t]
\caption{Symmetrically coupled molecular wires for which at zero bias the
HOMO-LUMO gap is centred at the Fermi level of the contacts and
$\epsilon_{contacts} = \epsilon_a = \epsilon_b$.
   a)Source and drain electrochemical potentials $\mu_S$ and $\mu_D$  and
energies of the bonding ($O$) and antibonding ($O^*$) orbitals as a
function of bias  voltage, for
$\beta_{-1,a} =
\beta_{1,b} = -1.0$ eV.
   b)Probability for the transmission of an electron through the molecular
wire as a function of electron energy, for $V_{bias} = 0$ and
$\beta_{-1,a} =
\beta_{1,b} = -1.0$ eV. c)Occupations of the molecular bonding and
antibonding orbitals, for $\beta_{-1,a} = \beta_{1,b} = -1.0$ eV, d)for
$\beta_{-1,a} = \beta_{1,b} = -0.2$ eV. e)Total integrated photon
emission, as a function of bias voltage, for (i)$\beta_{-1,a} = \beta_{1,b}
= -0.2$ eV and (ii)$\beta_{-1,a} = \beta_{1,b} = -1.0$ eV. f)Emission
spectra for various different bias voltages ($\beta_{-1,a} =
\beta_{1,b} = -1.0$ eV).}
\label{fig2}
\end{figure}

\begin{figure}[!t]
\caption{The case of $E_F$ close to either the HOMO or LUMO at zero bias.
$\beta_{-1,a} =
\beta_{1,b} = -1.0$ eV for the entire figure. a)$E_F$ is close to  HOMO at
zero bias. Energies of the bonding ($O$) and antibonding ($O^*$) orbitals
as a function of bias voltage. b)Energies of $O$ and $O^*$, for the case
of $E_F$ close to LUMO at zero bias. c)Occupations of $O$ and $O^*$, for
the situation in (a).  d)Occupations of $O$ and $O^*$, for the situation
in (b). e)Total photon emission rate for the situation in (a) (emission
rate for situation in (b) is very similar). f) Emission spectrum for the
situation in (a) for $V_{bias}=6$ V (Similar spectrum for situation in
(b)).}
\label{fig3}
\end{figure}

\begin{figure}[!t]
\caption{The energies of $O$ and $O^*$ as a function of bias voltage, for
the case of a HOMO LUMO gap that is centred at the zero bias Fermi level
of the contacts, with highly asymmetric contact couplings
($\beta_{-1,a}=-0.1$ eV, $\beta_{1,b}=-1.0$ eV).}
\label{fig4}
\end{figure}

\begin{figure}[!t]
\caption{The case of a HOMO-LUMO gap that is centred at the Fermi level of
the contacts zero bias, for various asymmetric  contact couplings.
a)$\beta_{-1,a}=-0.6$ eV,
$\beta_{1,b}=-1.0$ eV. Energies of the bonding ($O$) and antibonding
($O^*$) orbitals as a  function of bias voltage. b)Occupations of $O$ and
$O^*$, corresponding to the situation in (a). c)Total photon emission
rates, for various values of $\beta_{-1,a}$. $\beta_{1,b}=-1.0$ eV in all
cases. d)Current for various values of $\beta_{-1,a}$. $\beta_{1,b}=-1.0$
eV in all cases. e)Photon emission (solid line) and current (dotted line)
as a function of source contact coupling $\beta_{-1,a}$, for
$\beta_{1,b}=-1.0$ V,
$V_{bias}=6$ V. Vertical scales are arbitrary. f)Total photon  emission
vs. current, for
$\beta_{1,b}=-1.0$ V, $V_{bias}=6$ V.}
\label{fig5}
\end{figure}

\begin{figure}[!t]
\caption{a)The energies of $O$ and $O^*$ as a function of bias voltage,
for the case of $O$ located at $E_F$ at zero bias, with highly asymmetric
contact couplings ($\beta_{-1,a}=-0.1$ eV, $\beta_{1,b}=-1.0$ eV.)
b)Energies of $O$ and $O^*$ for reversed couplings ($\beta_{-1,a}=-1.0$
eV, $\beta_{1,b}=-0.1$ eV). c)Total photon emission, corresponding to the
situation in (a). d)Emission spectrum, corresponding to situation in (a),
for
$V_{bias}=6$ V. e)Energies of $O$ and $O^*$, for the case of $O^*$
located at $E_F$ at zero bias, with $\beta_{-1,a}=-0.1$ eV,
$\beta_{1,b}=-1.0$ eV.  f)Energies of $O$ and
$O^*$, for reversed couplings.}
\label{fig6}
\end{figure}


\begin{references}
\bibitem{exciton}The energies of the emitted photons  may in practice also
be influenced by excitonic correlation effects that are beyond the scope
of the present model. See, for example, Ref. \onlinecite{exciton1}.
\end{references}
\end{document}